\documentclass[letterpaper, 10 pt, conference]{ieeeconf}  

\IEEEoverridecommandlockouts                              

\usepackage{graphics} 
\usepackage{epsfig} 
\usepackage{mathptmx} 
\usepackage{amsmath} 
\usepackage{amssymb}  
\usepackage{algorithm}
\usepackage{algpseudocode}
\usepackage{algorithmicx}
\usepackage{cite}
\DeclareMathOperator*{\argmax}{arg\,max}

\title{\LARGE \bf
Trajectory Optimization for Spatial Microstructure Control in Electron Beam Metal Additive Manufacturing
}

\author{Mikhail Khrenov$^{1}$, Moon Tan$^{1}$, Lauren Fitzwater$^{2}$, Michelle Hobdari$^{1}$, and Sneha Prabha Narra$^{1}$
\thanks{$^{1}$Mikhail Khrenov, Moon Tan, Michelle Hobdari, and Sneha Prabha Narra are with the Department of Mechanical Engineering, Carnegie Mellon University,  Pittsburgh, PA, USA {\tt\small mkhrenov@cmu.edu, yuetnint@andrew.cmu.edu, mhobdari@andrew.cmu.edu, snarra@andrew.cmu.edu (corresponding author)}} 
\thanks{$^{2}$Lauren Fitzwater is with the Department of Materials Science and Engineering, Carnegie Mellon University, Pittsburgh, PA, USA {\tt\small lgf@andrew.cmu.edu}}%
}

\begin{document}

\maketitle
\thispagestyle{empty}
\pagestyle{empty}

\begin{abstract}
Metal additive manufacturing (AM) opens the possibility for spatial control of as-fabricated microstructure and properties. However, since the solid state diffusional transformations that drive microstructure outcomes are governed by nonlinear ODEs in terms of temperature, which is itself governed by PDEs over the entire part domain, solving for the system inputs needed to achieve desired microstructure distributions has proven difficult. In this work, we present a trajectory optimization approach for spatial control of microstructure in metal AM, which we demonstrate by controlling the hardness of a low-alloy steel in electron beam powder bed fusion (EB-PBF). To this end, we present models for thermal and microstructural dynamics. Next, we use experimental data to identify the parameters of the microstructure transformation dynamics. We then pose spatial microstructure control as a finite-horizon optimal control problem. The optimal power field trajectory is computed using an augmented Lagrangian differential dynamic programming (AL-DDP) method with GPU acceleration. The resulting time-varying power fields are then realized on an EB-PBF machine through an approximation scheme. Measurements of the resultant hardness shows that the optimized power field trajectory is able to closely produce the desired hardness distribution.
\end{abstract}

\section{Introduction}
Solid state diffusional transformations such as martensite tempering and precipitation hardening are dominant contributors to final strength in materials including steels, nickel superalloys, and certain aluminum alloys. While traditional metallurgy has developed methods to perform bulk control of such transformations through heat treatment, when temperature can be controlled directly as a single variable (e.g. in a furnace), the ability of additive manufacturing (AM) processes to vary power input with space has opened the possibility and challenge of location-specific control of microstructure and resulting material properties.

However, achieving this spatial control has proven nontrivial, as solid-state transformations at any point in a part are governed by dynamics which are highly nonlinear in the local temperature. The temperature itself, governed by a PDE, is driven by the process spatial power inputs, and both the temperature and power inputs are subject to equality and inequality constraints. Further, spatial discretization of the system can yield very high-dimensional state and input spaces, necessitating efficient solvers.

As a result of these difficulties, there has been limited success in spatial control of solid state transformations. The greatest successes in creating spatially varying microstructures in AM have focused on thermal conditions at solidification, whether through feedback control of cooling rates \cite{nair_effect_2024} or heuristic-based discrete planning \cite{plotkowski_stochastic_2021, dehoff_site_2015, NARRA2018160}, rather than solid state diffusional transformations. Though some work has attempted to achieve control of solid state transformations, it has sought to achieve this approximately by targeting the time spent within a certain temperature range, rather than directly modeling or making use of the microstructure dynamics \cite{MOZAFFAR2023103337}.

In this work, we present an approach for spatial control of microstructure and properties governed by diffusional transformations via augmented Lagrangian differential dynamic programming (AL-DDP) numerical trajectory optimization. We formulate the continuous and discrete time thermal and microstructure dynamics in Section \ref{sec:system-model}, experimentally identify the diffusional parameters in Section \ref{sec:system-identification}, formulate the optimal control problem in Section \ref{sec:optimal-control-problem}, and solve for optimal power field trajectories using a custom GPU accelerated trajectory optimizer and approximate the power-field solution via beam motion in Section \ref{sec:numerical-solution}. We experimentally demonstrate our approach through location specific control of hardness in a low-alloy steel sample processed using the EB-PBF process in Section \ref{sec:experimental-validation}. Finally, we discuss potential extensions, improvements, and insights in Section \ref{sec:conclusions}.

\section{System Model}
\label{sec:system-model}
The state of the AM part to be controlled is comprised of two scalar fields: the temperature, $T$, driven by the input power field $P$, and the extent of phase transformation $y$, with the former governed by a linear PDE and the latter a nonlinear ODE. We approach this problem through first discretizing in space, and then solving. The following subsections describe the modeling approach taken for each.


\subsection{Thermal Dynamics}
\label{subsec:thermal-model}
Temperature evolution in a solid is governed by the heat equation arising from Fourier's law, 
\begin{equation}
    \label{eq:heat-equation}
    \frac{\partial T}{\partial t} = \alpha \nabla^2 T + \frac{1}{\rho c_p} P
\end{equation}
a linear partial differential equation, where $\alpha$ is the thermal diffusivity $\frac{k}{\rho c_p}$, $k$ is the thermal conductivity, $\rho$ is the density, $c_p$ is the specific heat capacity, and $P$ is the input power field. In the case of EB-PBF, due to the fact that beam scan speeds can exceed the critical speed for thermal diffusion, power can be modeled as being applied over the top surface as a near arbitrary nonnegative field \cite{ADDOPT, Wood2023Theory}. 

As previously discussed in \cite{ADDOPT}, by applying a finite-volume spatial discretization with Dirichlet boundary conditions on the substrate, (\ref{eq:heat-equation}) can be transformed into an ordinary differential equation for the vector of finite volume temperatures, $\mathbf{T}$. If material properties are held to be temperature independent, the resulting ODE is affine and may be written as
\begin{equation}
    \label{eq:thermal-ode}
    \frac{d \mathbf{T}}{dt} = A \mathbf{T} + \mathbf{e} +  B \mathbf{u}
\end{equation}
Where for a uniform grid with discretization length $l$, $A = -(\frac{\alpha}{l^2}L + \frac{\alpha}{l^4} A_{0} + \frac{h}{\rho c_p l^3} A_{\infty})$, $L$ being the graph Laplacian of the finite volume mesh, $A_{0}$ being the diagonal matrix of the contact area between each volume and the substrate, $A_{\infty}$ being the diagonal matrix of the contact area between each volume and the atmosphere, and $h$ being the convective heat transfer coefficient \cite{ADDOPT}. Since electron beams must be used in a vacuum, we take $h=0$ for this work. $\mathbf{e}$ is the vector of exogenous inputs from the substrate and environment, $\mathbf{e} = (\frac{\alpha}{l^4}A_{0}T_{0} + \frac{h}{\rho c_p l^3}A_{\infty}T_{\infty})\mathbf{1}$, for substrate temperature $T_0$ and ambient temperature $T_{\infty}$. The input matrix $B$ maps from the magnitude of a Gaussian beam ($\mathbf{u}$) centered at each finite volume on the top surface of the part ($\mathcal{S}$), to the power field induced rate of temperature increase experienced by a volume, i.e.
\begin{equation}
    \label{eq:power-mapping}
    B_{ij} = \left\{ \begin{array}{ll}
        \frac{1}{\rho c_p l} \frac{1}{2\pi \sigma^2} \exp\left(-\frac{|\mathbf{r}_i - \mathbf{r}_j|^2}{2\sigma^2}\right) &  i \in \mathcal{S}\\
        0 & \text{else} 
    \end{array}\right.
\end{equation}
Since these dynamics are affine and time invariant, and assuming a 0$^{\text{th}}$ order hold on the input power field, they may be exactly discretized using the matrix exponential $\exp(A\Delta t)$ into a linear difference equation,
\begin{equation}
    \label{eq:thermal-discrete}
    \mathbf{T}_{k+1} = A_d \mathbf{T}_k + \mathbf{e}_d + B_d \mathbf{u}_k
\end{equation}

\subsection{Microstructure Transformation Dynamics}
\label{subsec:transformation-model}
The transformed phase fraction in solid state diffusional transformations such as the tempering of martensite in steels and precipitation strengthening of nickel superalloys and certain aluminum alloys for isothermal conditions are governed by the Johnson-Mehl-Avrami-Kolmogorov (JMAK) equation 
\begin{equation}
    \label{eq:avrami}
    y_f = 1 - \exp[-(\dot{G}(T) t)^n]
\end{equation}
Where $y_f$ is the final fraction of transformed phase, $n$ is the Avrami exponent, corresponding to the nature of nucleation and growth in the transformation, and $\dot{G}$ is the temperature dependent rate of transformation given by an Arrhenius reaction rate,
\begin{equation}
    \label{eq:arrhenius}
    \dot{G}(T) = A \exp\left(\frac{-E}{RT}\right)
\end{equation}
With pre-exponential factor $A$, activation energy $E$, and the universal gas constant $R = 8.3145 ~J \cdot \text{mol}^{-1} \cdot K^{-1}$. If the Avrami exponent $n$ is independent of temperature, then after applying the change of variables $\Tilde{y} = -\ln(1 - y)$ differentiating (\ref{eq:avrami}) yields the ODE \cite{todinov_alternative_1998}
\begin{equation}
    \label{eq:avrami-transformed}
    \frac{d \Tilde{y}}{dt} = n \dot{G}(T) \Tilde{y}^{1 - \frac{1}{n}}
\end{equation}
Assuming a small isothermal step, integrating (\ref{eq:avrami-transformed}) gives
\begin{equation}
    \label{eq:tilde-integration}
    \Tilde{y}_{k+1} = \left(\Tilde{y}_k^{1/n} + \dot{G}(T)\Delta t \right)^n
\end{equation}
These dynamics, as a result of (\ref{eq:arrhenius}), remain highly nonlinear in temperature. By taking an additional logarithm, $\hat{y} = \ln(\Tilde{y}) = \ln(-\ln(1-y))$, (\ref{eq:tilde-integration}) becomes
\begin{equation}
    \label{eq:hat-integration}
    \hat{y}_{k+1} = n\cdot \ln\left( \exp\left(\frac{\hat{y}}{n}\right) + \dot{G}(T)\Delta t\right)
\end{equation}
This transformed form has the advantage of removing the exponential temperature dependence.


\section{System Identification}
\label{sec:system-identification}
While the material properties such as thermal conductivity, density, etc. needed for the thermal dynamics may be found in literature or via CALPHAD (shown here in Table \ref{tab:properties}), in order to arrive at usable trajectories values for the parameters of the microstructure transformation dynamics, namely $n$, $A$, and $E$, are needed for the material in question.

\begin{table}[b]
    \centering
    \caption{Material (ER70S-6 low alloy steel) and machine (Freemelt ONE) parameters used for experiments.}
    \begin{tabular}{|c|c|c|c|}
        \hline
        Symbol & Property & Value & Units \\
        \hline
        $k$ & Thermal Conductivity & $31.1$ & $W \cdot m^{-1} \cdot K^{-1}$ \\
        $\rho$ & Density & $7269.0$ & $kg \cdot m^{-3}$\\
        $c_p$ & Specific Heat Capacity & $720.0$ & $J \cdot kg^{-1} \cdot K^{-1}$\\
        $\sigma$ & Beam Radius & $888.9$ & $\mu m$\\
        $P_{set}$ & Beam Power & $3000.0$ & $W$\\
        \hline
    \end{tabular}
    \label{tab:properties}
\end{table}


In order to identify these quantities for the tempering transformation, 16 samples were fabricated from the target material (ER70S-6 low alloy steel). These samples were quenched in order to achieve a largely martensitic structure and high hardness. After their post-quench hardness was measured by micro-indentation at four locations, the samples were subjected to heat treatment under a variety of temperatures and durations via resistive heating in a GLEEBLE thermomechanical simulator. Thermocouples attached along the length of the samples at the measurement locations recorded temperature measurements at 500 Hz throughout the treatments. Finally, hardness values were re-measured after heat treatment for all locations.

\begin{figure}[t]
    \centering
    \includegraphics[width=1.0\linewidth]{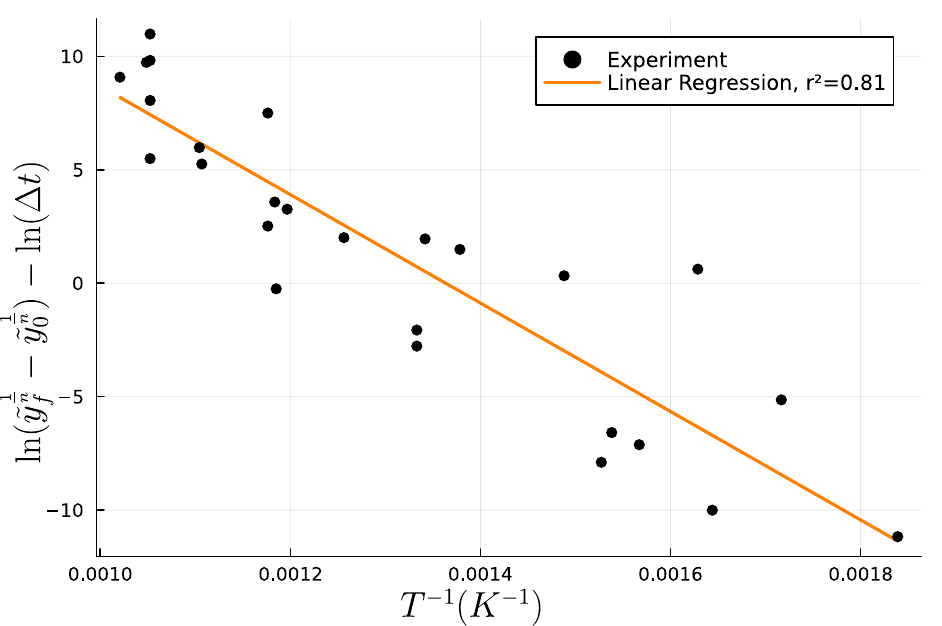}
    \caption{Fitted data resulting from a linear least squares regression with isothermal assumption and $n=0.05$.}
    \label{fig:sysid-fit-linear}
\end{figure}

This data was processed to obtain parameters in a two step process. First, a linear regression was used. For a fixed temperature, rearranging (\ref{eq:tilde-integration}) gives
\begin{equation}
    \Tilde{y}_{k+1}^{\frac{1}{n}} - \Tilde{y}_{k}^{\frac{1}{n}} = Ae^{\frac{-E}{RT}} \Delta t\\
\end{equation}
And by taking additional logarithm,
\begin{equation}
        \ln(\Tilde{y}_{k+1}^{\frac{1}{n}} - \Tilde{y}_{k}^{\frac{1}{n}}) - \ln(\Delta t) = \ln(A) - \frac{E}{RT}
\end{equation}
We arrive at an equation that is linear in the log of the pre-exponential factor and the activation energy. For each measurement site, the fraction of transformation was obtained by assuming a linear relationship between the extent of tempering and hardness, i.e.
\begin{equation}
    \label{eq:linear-hardness}
    HV = HV_{\max} - y \cdot (HV_{\max} - HV_{\min})
\end{equation}
The temperature was taken to be the median temperature experienced over the duration of the experiment, the time was measured as the active on time, and $n$ was assumed to be 0.05 based on prior literature on steels \cite{costa_tempering_2003}. The resulting linear fit is shown in Fig. \ref{fig:sysid-fit-linear}.

\begin{figure}[t]
    \centering
    \includegraphics[width=1.0\linewidth]{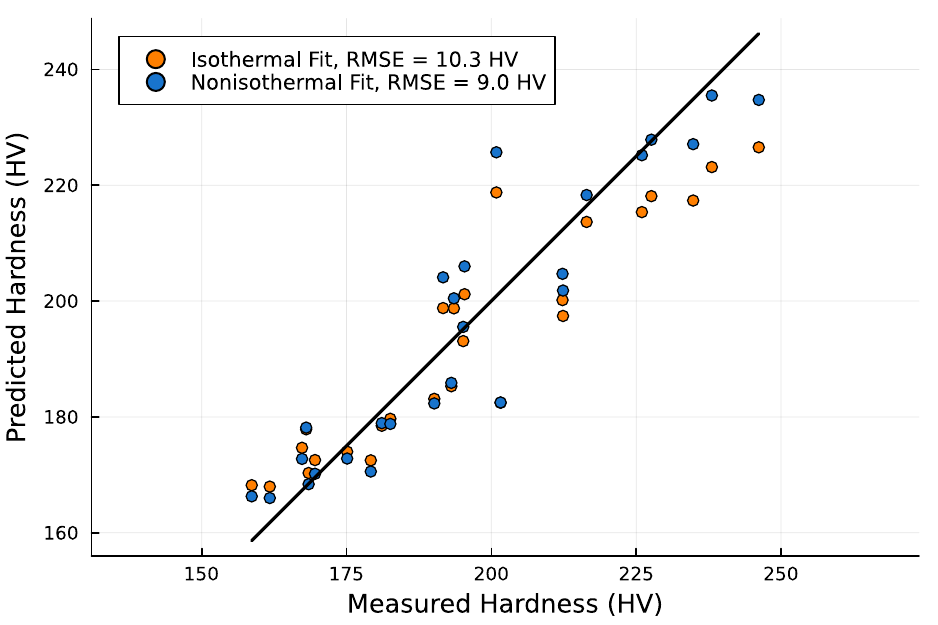}
    \caption{Correspondence between predicted and measured hardness for the samples using both the isothermal fit with linear regression, and nonlinear regression for  nonisothermal data, plotted against the perfect prediction line.}
    \label{fig:sysid-fit-comparison}
\end{figure}

The parameters fitted using this linear regression were then refined by taking them as initial values in a nonlinear least squares optimization with loss $\mathcal{L} = |\hat{y}^p(A, n, E) - \hat{y}^m|^2$ where $\hat{y}^m$ is the measured and $\hat{y}^p$ the predicted transformation given by integrating the recorded thermal data
\begin{equation}
    \hat{y}_i^p(A, n, E) = \int_{t_0}^{t_f} \frac{d\hat{y}}{dt}(A, n, E) dt
\end{equation}
This integral was approximated by taking each measurement to be an isothermal step as in (\ref{eq:hat-integration}) and back-propagating the gradient and hessian. The resulting nonlinear least squares was solved with the IPOPT interior-point Newton-based solver \cite{wachter_implementation_2006}. In order to avoid an ill-conditioned Hessian, variables were transformed to $\theta = [n\ln(A)~ 1/n~ nE]^T$. A comparison of the isothermal linear and nonisothermal nonlinear fits is shown in Fig. \ref{fig:sysid-fit-comparison}. The calibrated $n$, $\ln(A)$, and $E$ are given in Table \ref{tab:avrami-params}.

\begin{table}[hb]
    \centering
    \caption{Avrami parameters, determined by linear and nonlinear regression from rod experiments.}
    \begin{tabular}{|c|c|c|c|}
        \hline
        Parameter& Isothermal Fit   & Nonisothermal Fit   &    Units \\
        \hline
        $\ln(A)$ & 32.583           & 38.005           & $\ln(s^{-1})$\\
        $n$      & 0.05             & 0.05159          & unitless\\
        $E$      & 198.68           & 240.24           &  $kJ \cdot \text{mol}^{-1} K^{-1}$\\
        \hline
    \end{tabular}
    \label{tab:avrami-params}
\end{table}

\section{Optimal Control Problem}
\label{sec:optimal-control-problem}
In order to achieve the desired final transformation distribution, we seek to solve the optimal control problem,
\begin{equation}
    \label{eq:optimal-control-continuous}
    \begin{split}
        \min_{\mathbf{x}(t), \mathbf{u}(t)} ~~&l_f(\hat{\mathbf{y}}(t_f)) + \int_{t_0}^{t_f}  l(\hat{\mathbf{y}}(t)) dt \\
          s.t.~& \dot{\mathbf{x}} = f(\mathbf{x}, \mathbf{u})\\
               & 1^T \mathbf{u}(t) = P_{set} ~t \in [t_0, t_b]\\
               & \mathbf{u}(t) \geq \mathbf{0} ~t \in [t_0, t_b]\\
               & \mathbf{u}(t) = \mathbf{0} ~t \in [t_b, t_f]\\
               & \mathbf{T}(t) \leq \mathbf{T}_{max}
    \end{split}
\end{equation}
The dynamics $f(x,u)$ are as described in Section \ref{sec:system-model}. Due to the nature of the EB-PBF hardware, total power over the target area is constrained to either sum to a fixed value ($P_{set}$) during the exposure window $[t_0, t_b]$ or be uniformly 0 afterwards. The total time horizon of the optimization is extended beyond the beam exposure time to $t_f$ since transformations continue as the material is cooling. Temperature is bounded above in order to prevent a different microstructure transformation (re-austenitization, not modeled in this work) from occurring. The stage and final costs are the weighted squared error between the resultant and target distributions, $
    l_f = \frac{1}{2}(\hat{\mathbf{y}}(t_f) - \bar{\hat{\mathbf{y}}})^T Q_f (\hat{\mathbf{y}}(t_f) - \bar{\hat{\mathbf{y}}})$ and $
    l = \frac{1}{2}(\hat{\mathbf{y}}(t) - \bar{\hat{\mathbf{y}}})^T Q (\hat{\mathbf{y}}(t) - \bar{\hat{\mathbf{y}}})$
for diagonal $Q$ and $Q_f$, where the diagonal values of $Q$ are much smaller than $Q_f$. This problem can be discretized in time over $N$ steps,
\begin{equation}
    \label{eq:optimal-control-discrete}
    \begin{split}
        \min_{\mathbf{x}_{1:N}, \mathbf{u}_{1:N}} ~~&l_f(\hat{\mathbf{y}}(t_f)) + \sum_{k=1}^{N}  l(\hat{\mathbf{y}}_k) \Delta t \\
          s.t.~& \mathbf{x}_{k+1} = f_d(\mathbf{x}_k, \mathbf{u}_k)\\
               & \mathbf{1}^T \mathbf{u}_k = P_{set} ~k \in [1, N_b]\\
               & \mathbf{u}_k \geq \mathbf{0} ~k \in [1, N_b]\\
               & \mathbf{u}_k = \mathbf{0} ~k \in [N_b + 1, N]\\
               & \mathbf{T}_k \leq \mathbf{T}_{max}
    \end{split}
\end{equation}
The thermal and microstructural dynamics are discretized using (\ref{eq:thermal-discrete}) and (\ref{eq:hat-integration}) respectively, with stage costs integrated.

\section{Numerical Solution via AL-DDP}
\label{sec:numerical-solution}
The optimal control problem from (\ref{eq:optimal-control-discrete}) can be solved using augmented Lagrangian differential dynamic programming (AL-DDP). Traditional DDP solves optimal control problems by iteratively rolling out the nonlinear dynamics in a forward pass, then sequentially calculating differential updates to the input at each time step in a backward pass \cite{DDP}. These updates arise from the Bellman equation, which states that the optimal remaining cost (cost-to-go) for a given timestep and state, $V(\mathbf{x}_k)$, is given by the recursion 
\begin{equation}
    \label{eq:bellman}
    V(\mathbf{x}_k) = \min_{\mathbf{u}_k}  ~l(\mathbf{x}_k, \mathbf{u}_k) \Delta t + V(f_d(\mathbf{x}_k, \mathbf{u}_k))
\end{equation}
By approximating $V(\mathbf{x}_k)$ and the action-value function $l(\mathbf{x}_k, \mathbf{u}_k) \Delta t + V(f_d(\mathbf{x}_k, \mathbf{u}_k))$ using a second-order Taylor expansion about a given forward simulated state and inputs $\mathbf{x}_k$ and $\mathbf{u}_k$, the minimization in (\ref{eq:bellman}) can be made into an unconstrained quadratic program (QP) in terms of the $\Delta \mathbf{u}_k$, which has a closed form, establishing a recursive relation. The forward and backward steps are repeated until the input trajectory converges. While traditional DDP solves unconstrained optimal control problems, support for constraints can be introduced by the the use of an augmented Lagrangian \cite{AL-DDP} with a penalty term in the stage cost. An outer loop increases the penalty weight, until constraints are satisfied to desired tolerance.

Compared to simultaneous methods such as direct collocation (DIRCOL), DDP offers substantially better scaling for high dimensional systems. Since simultaneous methods combine all $m$ dimensional inputs and  $n$ dimensional states for all $N$ time steps into a single decision vector, to use a second order update they must effectively solve/invert a $(2n+m)\cdot N$ by $(2n+m)\cdot N$  matrix, which scales cubically in general, and so one step will scale as $O(((2n+m)\cdot N)^3)$. In contrast, since DDP only solves QPs in terms of the inputs at a single time step, it must only invert an $m$ by $m$ matrix at each time step. As such, a single forward and backward iteration of DDP scales as $O(m^3\cdot N)$. 

DDP is closely related to iLQR \cite{iLQR}, which uses a first (rather than second) order approximation of the dynamics. While iLQR is widely used and AL-iLQR implementations are available \cite{ALTRO}, we have found that due to the strong nonlinearities in the microstructure dynamics this problem benefits substantially from including the higher order derivative terms. Due to the very high dimensionality of the state and input spaces, we implemented AL-DDP with GPU acceleration within our AM optimal control (ADDOPT) system \cite{ADDOPT}.

\begin{figure}[t]
    \centering
    \includegraphics[width=0.75\linewidth]{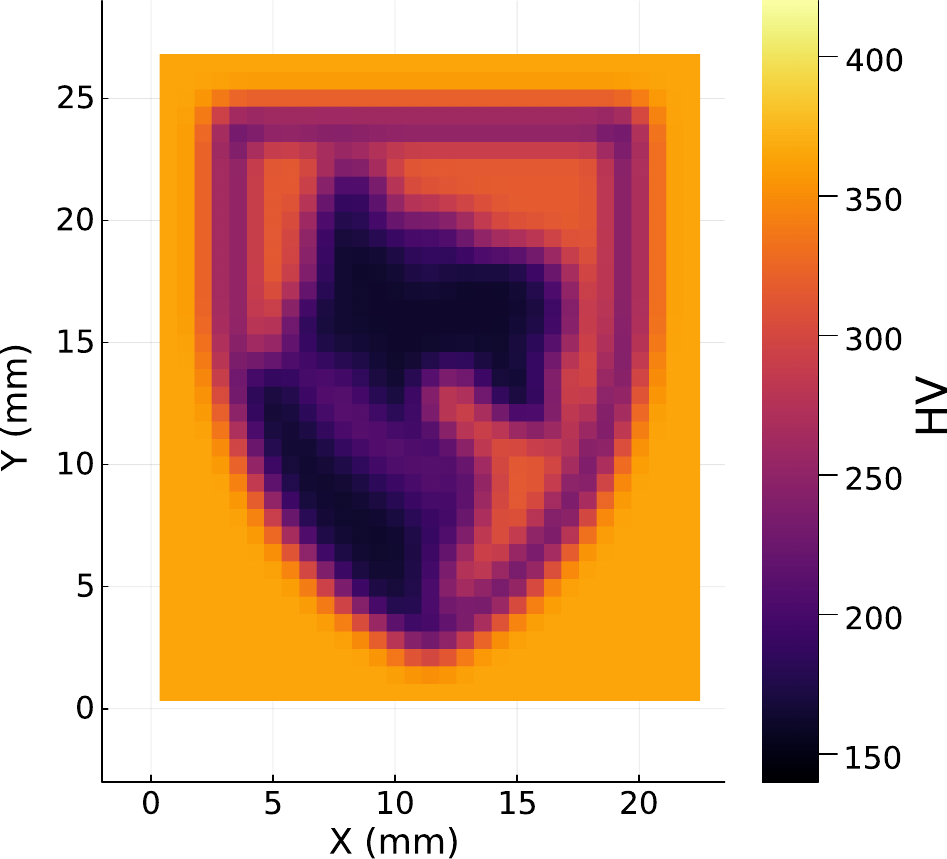}
    \caption{Target hardness distribution for demonstration, a downsampled version of Carnegie Mellon University's ``Scotty'' mascot. The hardness range is set by the empirically determined maximum hardness of 420 HV (fully martensitic) and minimum hardness of 135 HV (fully tempered).}
    \label{fig:target-geometry}
\end{figure}

\begin{figure}[t]
    \centering
    \vspace{.075in}
    \includegraphics[width=1.0\linewidth]{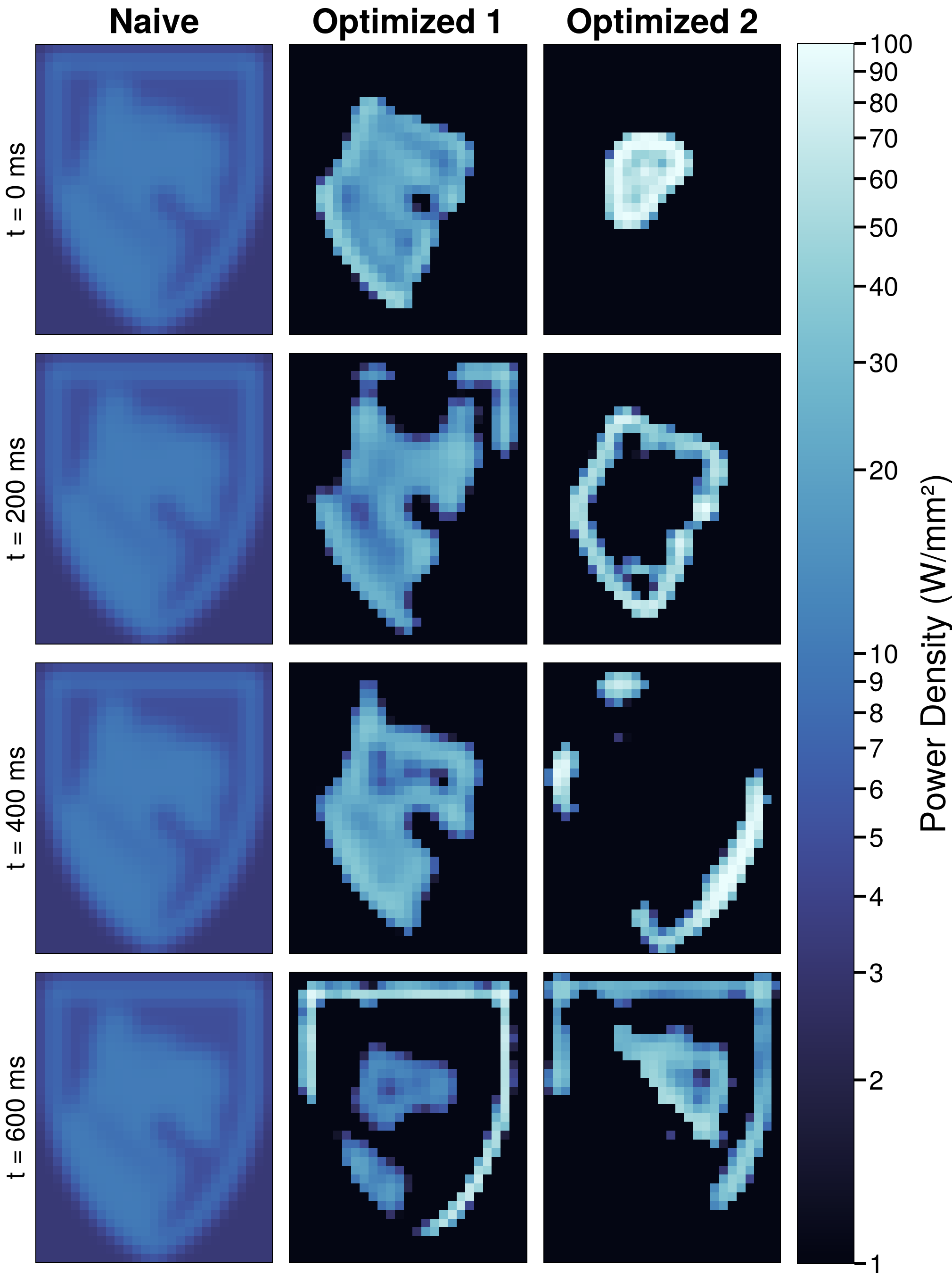}
    \caption{Naive and optimized power fields over time. Optimized 1 is the optimal power field for the parameters identified in Section \ref{sec:system-identification}, while Optimized 2 is the same for the updated parameters from Section \ref{sec:experimental-validation}.}
    \label{fig:opt-power-fields}
\end{figure}

\begin{figure}[t]
    \centering
    \vspace{.05in}
    \includegraphics[width=1.0\linewidth]{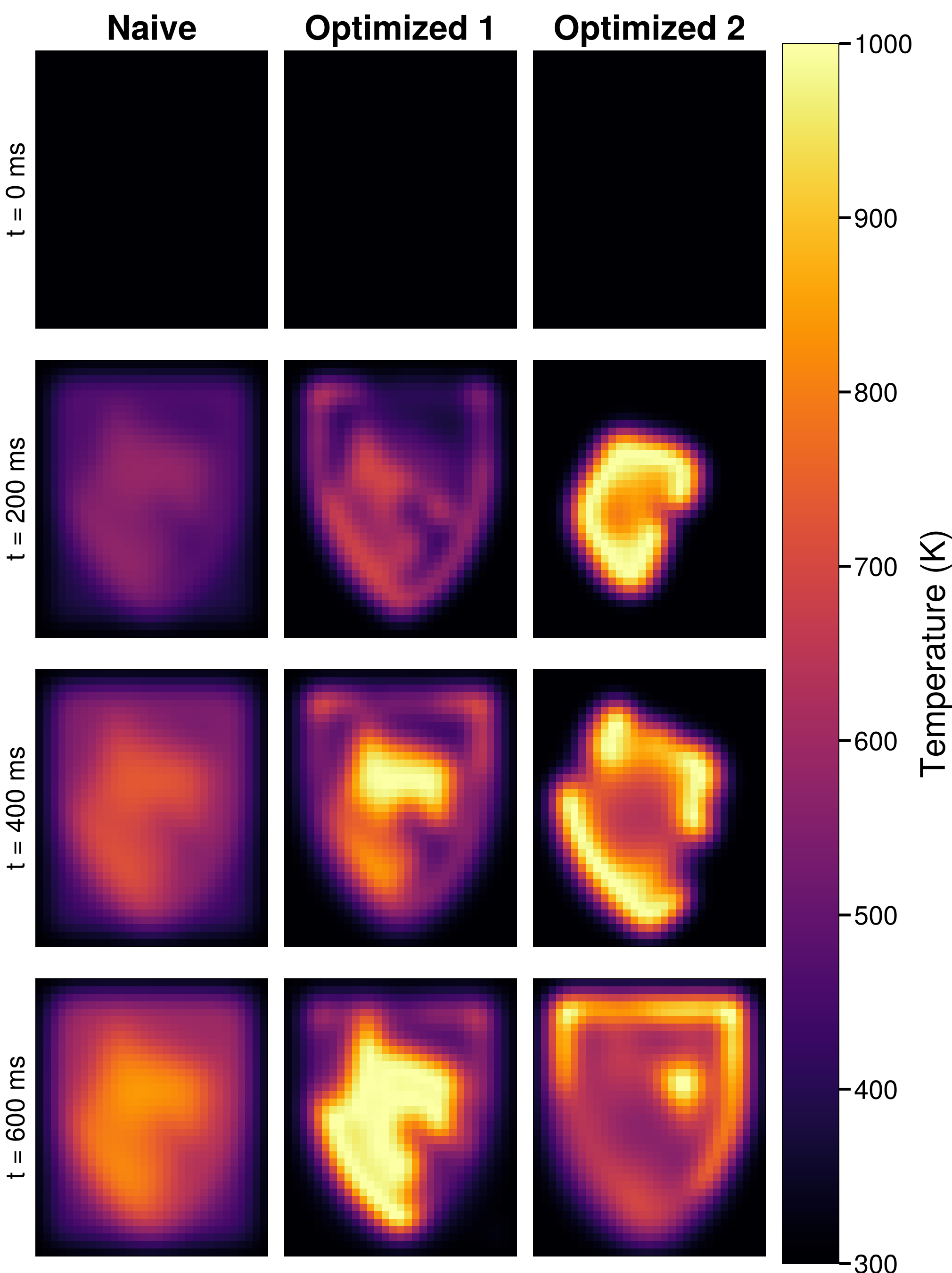}
    \caption{Temperature fields resulting from the naive and two optimized power field trajectories over time.}
    \label{fig:opt-temp-fields}
\end{figure}

For demonstration, we targeted the hardness distribution shown in Fig. \ref{fig:target-geometry} over a 22 mm by 26 mm area. We discretized the domain into 4588 cubic volumes with 0.7 mm side lengths, for a total of 9176 state variables, with the top surface power field parameterized by 891 Gaussians. The system was discretized in time for 27 time steps of 50 ms, with the beam exposure enabled for the first 13 and the rest reserved for cooling. Solving this demonstration case took approximately 10 minutes of wall clock time using a single NVIDIA RTX 4090 GPU. As a naive baseline for comparison, we take a time-invariant power field proportional to the desired extent of transformation. The naive and optimized power field trajectories and simulated resulting thermal histories are shown in Fig. \ref{fig:opt-power-fields} and Fig. \ref{fig:opt-temp-fields} respectively.

\begin{algorithm}[b]
\caption{Greedy beam approximation of the power field}\label{alg:approximation}
\begin{algorithmic}
\Procedure{ApproxField}{$\mathbf{u}_{1:N}$, $\Delta t_{opt}$, $\Delta t_{dwell}$, $P_{set}$, $r_{1:n}$}
    \State $b \gets [~]$

    \For{$k \in {1 \dots N}$}
        \State $t \gets 0$
        \While{$t < \Delta t_{opt}$}
            \State $i \gets \argmax_i \mathbf{u}_k[i]$
            \State $\mathbf{u}_k[i] \gets \mathbf{u}_k[i] - P_{set} \frac{\Delta t_{dwell}}{\Delta t_{opt}}$
            \State \textbf{append} $r[i]$ \textbf{to} $b$
            \State $t \gets t + \Delta t_{dwell}$
        \EndWhile
    \EndFor
    \State \Return $b$
\EndProcedure
\end{algorithmic}
\end{algorithm}

In order to produce the target time-varying power fields using a beam, we made use of a simple time-averaging scheme reminiscent of pulse width modulation. In essence, we select beam motions by greedily seeking to reproduce the solution from Section \ref{sec:numerical-solution} by averaging the beam's total power input to each surface volume location over the time of an optimizer time step.
This procedure is detailed in Algorithm \ref{alg:approximation}, where $u_{1:N}$ is the sequence of power fields to be approximated, $\Delta t_{opt}$ is the duration of each power field (50 ms), $\Delta t_{dwell}$ is the dwell time for each beam motion (27 $\mu$s), $P_{set}$ is the beam power (3 kW), and $r_{1:N}$ is the array of coordinates corresponding to each power field Gaussian magnitude in $u$. The procedure returns $b$, a sequence of beam coordinates. Dwell times are chosen long enough such that beam motion time is negligible.

\section{Experimental Validation}
\label{sec:experimental-validation}
Microstructural control experiments were carried out on a Freemelt ONE EB-PBF system. A 60 kV, 50 mA beam was used, resulting in a total power of 3 kW. The experimental sample was prepared by fabricating and quenching a 100 mm diameter, 10 mm thick disc of ER70S-6 steel. The initial hardness was measured with 3.0 mm spaced microindents.

\begin{figure*}[t]
    \centering
    \vspace{.05in}
    \includegraphics[width=1.0\linewidth]{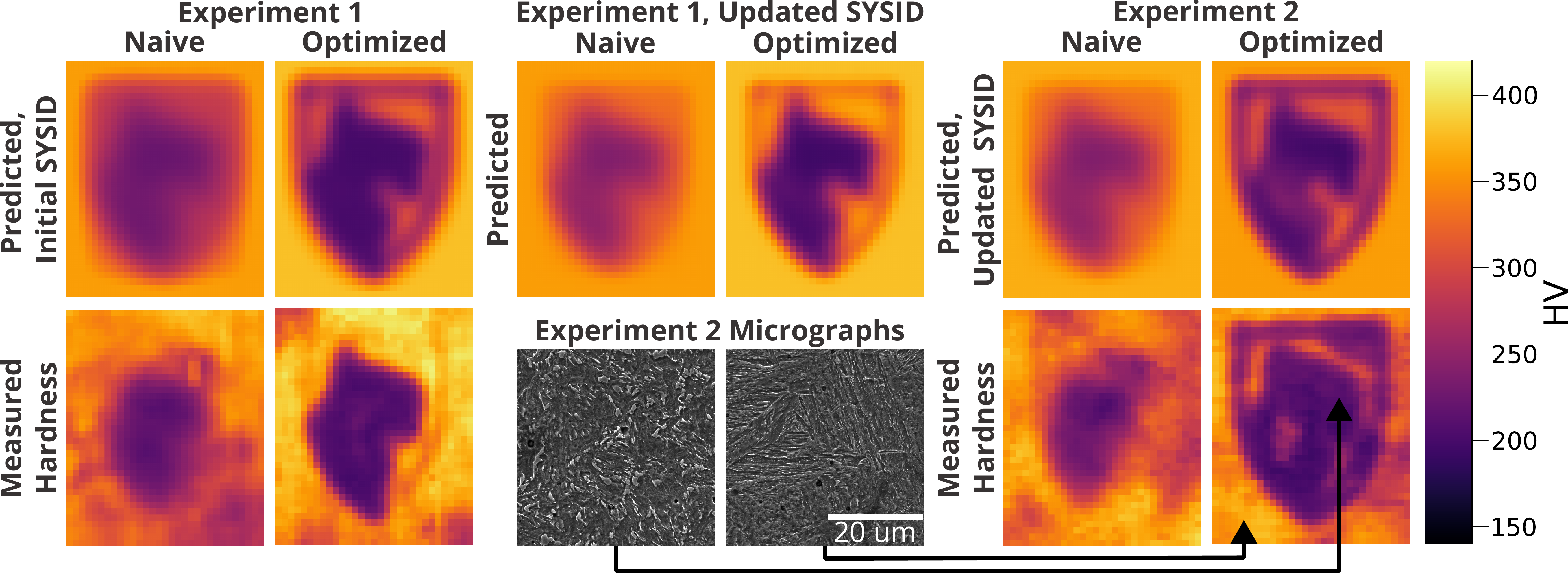}
    \caption{Left: Comparison of simulated and empirical final hardness measurements for Experiment 1. Center Top: Simulated final hardness measurements for the Experiment 1 power field trajectories after updating Avrami parameters using Experiment 1 measurements. Right: Comparison of simulated and empirical final hardness measurements for Experiment 2. Center Bottom: Secondary electron micrographs of two points on the final optimized sample.}
    \label{fig:results-comparison}
\end{figure*}

The naive and optimal beam trajectories were executed on the EB-PBF machine. The expected and measured hardnesses (measured with the same 0.7 mm spacing as the spatial discretization) are shown on the left of Fig. \ref{fig:results-comparison}.  While the optimized profile resulted in substantially closer reproduction of the target hardness distribution from Fig. \ref{fig:target-geometry} compared to the naive power field, it did not match the expectation from simulation, especially on the outer edges.

As such, the microstructure dynamics parameters were updated by running a second nonlinear least squares minimization, with the measured hardness distribution serving as the ground truth and the hardness predicted by rolling out the experimental power fields for a candidate parameter set. As in Section \ref{sec:system-identification}, an interior point solver was used, giving $\ln(A) = 8.845$, $n = 0.358$, and $E = 56.277 ~\frac{kJ}{ \text{mol}}$. The difference in values is likely due to the absence of low transformation samples in the initial experiments. The new simulation of the Experiment 1 power fields with updated parameters is shown in the center of Fig. \ref{fig:results-comparison}, showing a close fit to the measured hardness from Experiment 1. 

Using these updated microstructure dynamics parameters, the power field was then re-optimized, with the updated trajectory shown as Optimized 2 in Fig. \ref{fig:opt-power-fields} and Fig. \ref{fig:opt-temp-fields}. The new trajectory was approximated by beam motions and executed, yielding the measured hardness distributions, shown at the right of Fig. \ref{fig:results-comparison}. This yielded substantially better agreement between target, prediction, and measured hardness values, resulting in a 51.1\% reduction in RMSE compared to the naive strategy. SEM micrographs of several points on the final sample in Fig. \ref{fig:results-comparison} shows a more needle-like structure for harder regions than softer regions, consistent with martensite tempering.

\section{Conclusions}
\label{sec:conclusions}
We have demonstrated spatial control of microstructure and properties in EB-PBF AM process. We accomplished this by applying discretized models for the thermal and microstructural dynamics, identifying the model parameters, and solving an optimal control problem using AL-DDP. The optimized power fields were tested experimentally, yielding a strong agreement with the target hardness distribution.

This work opens the door to both practical and theoretical extensions. For instance, on the practical side, this approach could be applied to achieve selective control of hardness for improving machinability for hybrid manufacturing. On the theoretical side, more analysis is needed of the properties of systems governed by diffusional dynamics. We also believe there are still possible improvements in speed and efficiency of trajectory optimizers for AM systems. Better analysis is also needed to determine final state reachability, as well as real-time material parameter estimation methods to enable applications in large-scale manufacturing.



\section*{Acknowledgments}
The authors thank Professor P. Chris Pistorius and Pooja Maurya for help in quenching samples, and Delaynie McMillan for help in machining samples. The authors also thank Alexander Myers, William Frieden Templeton, and Nathan Wassermann for help in experimental setup.

This material is based upon work supported by the National Science Foundation Graduate Research Fellowship under Grant No. 2140739. Experimental work was supported by the Next Manufacturing Center, the Manufacturing Futures Institute, and the Carnegie Mellon University Office of Undergraduate Research and Scholar Development.  

A portion of this research was supported by the Army Research Laboratory under Cooperative Agreement Number W911NF-20-2-0175. The views and conclusions contained in this document are those of the authors and should not be interpreted as representing the official policies, either expressed or implied, of the Army Research Laboratory or the U.S. Government. The U.S. Government is authorized to reproduce and distribute reprints for Government purposes notwithstanding any copyright notation herein. 

\bibliographystyle{bib/IEEEtran}
\bibliography{bib/IEEEabrv,bib/IEEEfull,bib/IEEEbcpat,bib/bib}

\end{document}